\documentclass{appolb}
\usepackage{epsfig}
\begin{document}
% \eqsec  % uncomment this line to get equations numbered by (sec.num)
\title{Thermoelectric transport through a quantum dot coupled to a normal metal
and BCS superconductor}
\author{Mariusz Krawiec
\address{Institute of Physics, M. Curie-Sk\l odowska University, \\
pl. M. Curie-Sk\l odowskiej 1, 20-031 Lublin, Poland}
}
\maketitle

\begin{abstract}
I discuss thermoelectric properties of a quantum dot coupled to one normal and 
one superconducting lead in the presence of Kondo effect and Andreev 
scattering. I will focus on conductance, thermal conductance, thermopower and 
related quantities like thermoelectric figure of merit which is a direct 
measure of the usefulness of the system for applications and Wiedemann-Franz 
ratio which indicates if the system is in the Fermi liquid state. I will show 
that the superconductivity strongly modifies the thermal properties of the 
system. In particular, the thermopower is strongly enhanced near the 
superconducting transition temperature. Moreover, the Andreev reflections are 
suppressed due to strong on-dot Coulomb repulsion. The suppression of the 
Andreev reflections leads to a violation of the Wiedemann-Franz law and to a 
non-Fermi liquid ground state.
\end{abstract}

%%%%%%%%%%%%%%%%%%%%%%%%%%%%%%%%%%%%%%%%%%%%%%%%%%%%%%%%%%%%%%%%%%%%%%%%%%%%%%  

\section{Introduction}
Quantum dot (QD) devices provide a unique opportunity to study the interplay 
between different basic quantum phenomena. A prime example is the interaction 
of a localized spin (magnetic impurity) with surrounding itinerant electron 
spins \cite{Hewson}. This, known as Kondo effect, manifests itself as a 
quasiparticle resonance in local density of states at the Fermi level at low 
temperatures. In quantum dots coupled to normal electrodes, it leads to an 
increase of conductance at zero source-drain voltage. Originally, the Kondo
effect in quantum dots was predicted theoretically in the late 1980s 
\cite{Glazman}-\cite{Kawabata} and later demonstrated in a number of 
experiments \cite{Goldhaber}-\cite{Buitelaar}. 

Experiments have confirmed the validity of the theoretical picture but also 
discovered new phenomena. Those include observation of the Kondo resonance at 
non-zero source-drain voltage \cite{Schmid,Simmel}, absence of even-odd parity 
effects expected for these systems \cite{Schmid_1}, observation of 
singlet-triplet transition in magnetic field \cite{Sasaki}, splitting of the 
Kondo resonance due to the ferromagnetism in the leads \cite{Pasupathy} or 
interplay between magnetism (Kondo effect) and superconductivity in carbon 
nanotube quantum dots \cite{Buitelaar,Eichler}. 

When the quantum dot is connected to superconducting electrodes (S-QD-S), the 
low energy transport is mediated by the Andreev reflections, according to which 
an electron impinging on a normal metal-superconductor interface is reflected 
back as a hole and the Cooper pair is created in the superconductor 
\cite{Andreev}. This mechanism is known to play a crucial role in various 
hybrid mesoscopic superconducting devices \cite{Lambert}. In the S-QD-S system
the Andreev transport is strongly affected by the Kondo effect, leading to the
sign change of Josephson current \cite{Choi,Sellier}. The strong Coulomb
interaction prevents the tunneling of Cooper pairs into QD, and the electrons
tunnel via virtual processes. Thus the electron transport is strongly
suppressed. This argument is valid when the Kondo temperature $T_K$ is smaller
than the superconducting order parameter $\Delta$. On the other hand, when
$T_K > \Delta$, the Kondo effect is restored \cite{Avishai}. This picture has
been confirmed experimentally in carbon nanotube QD \cite{Buitelaar}.

In the present work I consider a slightly different setup with a quantum dot
coupled to one normal and one superconducting electrode (N-QD-S). This system
has extensively been studied both theoretically 
\cite{Fazio}-\cite{Donabidowicz_2} and experimentally \cite{Graber}. Early
theoretical works have predicted the enhancement of the conductance due to the
Andreev reflections \cite{Schwab,Kang}, while the others have predicted the
suppression of it \cite{Fazio,Clerk,Cuevas,MK_1}. Later on, it has been shown
that the behavior of the Andreev conductance depends on the model parameters 
and can be enhanced or suppressed \cite{Cuevas}. The experiment by Graber {\it
et al.} \cite{Graber} shows that the conductance of N-QD-S system is suppressed 
in the Kondo regime. 

The investigations of the N-QD-S systems have focused on electron transport 
only, and there is no study of thermoelectric properties. However, as is well 
known, the thermoelectric properties are the source of additional information 
to that obtained from other transport characteristics. Thermal properties 
(thermopower and thermal conductance) of strongly interacting quantum dot 
coupled to the normal and ferromagnetic leads have recently been investigated 
showing that thermopower is very sensitive and powerful tool to study the Kondo 
effect \cite{Boese}-\cite{MK_4}.

It is the purpose of the present work to study the thermoelectric properties of
the quantum dot coupled to one normal and one superconducting electrode. I will
focus on the electric and thermal conductance, thermopower and related 
quantities like thermoelectric figure of merit which is a direct measure of the
usefulness of the system for applications and Wiedemann-Franz ratio which
indicates if the system is in the Fermi liquid state. I will show that the
superconductivity strongly modifies the thermal properties of the system 
leading to an increase of thermopower at low temperatures and to non-Fermi 
liquid ground state.

Rest of the paper is organized as follows. In Sec. \ref{model} I present the
model and briefly discuss the approach. Results of calculations are presented
and discussed in Sec. \ref{results}. Summary and conclusions are given in 
Sec. \ref{conclusions}.

%%%%%%%%%%%%%%%%%%%%%%%%%%%%%%%%%%%%%%%%%%%%%%%%%%%%%%%%%%%%%%%%%%%%%%%%%%%%%%  

\section{\label{model} Formulation of the problem}

The system is described by single impurity Anderson model with very strong
on-dot Coulomb repulsion ($U \rightarrow \infty$). In this limit there might be
at most a single electron on the dot. To project out double occupied states it
is convenient to introduce slave boson representation \cite{LeGuillou,MK_5}, 
in which the real dot electron operator $d_{\sigma}$ is replaced by the product
of boson $b$ and fermion $f_{\sigma}$ operators ($d_{\sigma} = b^+ f_{\sigma}$)
subject to the constraint $b^+ b + \sum_{\sigma} f^+_{\sigma} f_{\sigma} = 1$.
The resulting Hamiltonian reads
\begin{eqnarray}
H = \sum_{\lambda {\bf k} \sigma} \epsilon_{\lambda {\bf k}}
    c^+_{\lambda {\bf k} \sigma} c_{\lambda {\bf k} \sigma} +
    \sum_{\bf k} \left(\Delta
    c^+_{S {\bf k} \uparrow} c^+_{S -{\bf k} \downarrow} + H. c. \right)
\nonumber \\
    + \varepsilon_d \sum_{\sigma} f^+_{\sigma} f_{\sigma} +
    \sum_{\lambda {\bf k} \sigma} \left(V_{\lambda {\bf k}}
    c^+_{\lambda {\bf k} \sigma} b^+ f_{\sigma} + H. c. \right),
\label{Hamilt}
\end{eqnarray}
where $\lambda = N$ ($S$) denotes normal (superconducting) electrode, 
$c^+_{\lambda {\bf k} \sigma}$ ($c_{\lambda {\bf k} \sigma}$) is the creation
(annihilation) operator for a conducting electron with the wave vector $\bf k$,
spin $\sigma$ in the lead $\lambda$, and $V_{\lambda {\bf k}}$ is the
hybridization parameter between localized electron on the dot with energy
$\varepsilon_d$ and conducting electron of energy $\epsilon_{\lambda {\bf k}}$
in the lead $\lambda$. $\Delta$ is the superconducting order parameter in the 
lead $S$, and its temperature dependence is chosen in the form: 
$\Delta(T) = \Delta_0 \sqrt{1 - (T/T_c)^2}$ in superconducting state 
($T \leq T_c$) and 0 otherwise. 

In order to calculate electric current $J_e$ and thermal flux $J_Q$ flowing 
from the normal electrode to the QD I follow standard derivation 
\cite{Haug,MK_1} and get
\begin{eqnarray}
 J_e = J^{sp}_e +J^A_e = 
      \frac{e}{h} \int d\omega \mathcal{T}^{sp}(\omega)
      \left[f(\omega - eV) - f(\omega) \right] 
 \nonumber \\ 
      + \frac{e}{h} \int d\omega \mathcal{T}^A(\omega)
      \left[f(\omega - eV) - f(\omega + eV) \right] ,
 \label{e_current}
\end{eqnarray}
\begin{eqnarray}
J_Q = J^{sp}_Q +J^A_Q = 
     \frac{1}{h} \int d\omega \mathcal{T}^{sp}(\omega) (\omega - eV)
     \left[f(\omega - eV) - f(\omega) \right] 
 \nonumber \\
     + \frac{1}{h} \int d\omega \mathcal{T}^A(\omega) (\omega - eV)
     \left[f(\omega - eV) - f(\omega + eV) \right] ,
\label{Q_current}
\end{eqnarray}
where the total current $J_e$ and thermal flux $J_Q$ are divided into two parts,
one associated with single particle processes ($sp$) and the other one
corresponding to the Andreev reflections ($A$). As usual, $f(\omega)$ stands
for the Fermi distribution function, and $eV = \mu_N - \mu_S$. 
$\mathcal{T}^{sp}(\omega)$ and $\mathcal{T}^A(\omega)$ are the transmittances 
associated with single particle tunneling and with Andreev reflections, 
respectively. The explicit form of $\mathcal{T}^{sp(A)}(\omega)$ and the 
details of calculations can be found in Ref. \cite{MK_1}.

In the linear regime, i.e. for small voltages $eV \rightarrow 0$ and small
temperature gradients $\delta T = T_N - T_S \rightarrow 0$, one defines the
electric conductance $G = -\frac{e^2}{T} L_{11}$, thermopower 
$S = -\frac{1}{eT} \frac{L_{12}}{L_{11}}$, and thermal conductance 
$\kappa = \frac{1}{T^2} \left(L_{22} - \frac{L^2_{12}}{L_{11}} \right)$. The
kinetic coefficients read
\begin{eqnarray}
L_{11} = \frac{T}{h} \int d\omega \mathcal{T}^{tot}(\omega)
\left(\frac{\partial f(\omega)}{\partial \omega} \right)_{T} ,
\label{L11}
\end{eqnarray}
\begin{eqnarray}
L_{12} = \frac{T^2}{h} \int d\omega \mathcal{T}^{tot}(\omega)
\left(\frac{\partial f(\omega)}{\partial T} \right)_{eV} ,
\label{L12}
\end{eqnarray}
\begin{eqnarray}
L_{22} = \frac{T^2}{h} \int d\omega \mathcal{T}^{tot}(\omega)
(\omega - eV)
\left(\frac{\partial f(\omega)}{\partial T} \right)_{eV} ,
\label{L22}
\end{eqnarray}
with equilibrium Fermi function $f(\omega)$ and total (single particle and
Andreev) transmittance $\mathcal{T}^{tot}(\omega)$. In numerical calculations 
constant bands of width 100 $\Gamma$ ($\Gamma = 2 \pi (V^2_S+V^2_N)$) have 
been assumed, and all the energies are measured in units of $\Gamma$.

%%%%%%%%%%%%%%%%%%%%%%%%%%%%%%%%%%%%%%%%%%%%%%%%%%%%%%%%%%%%%%%%%%%%%%%%%%%%%%  

\section{\label{results} Results and discussion}

\begin{figure}[h]
\begin{center}
\includegraphics[width=0.47\textwidth]{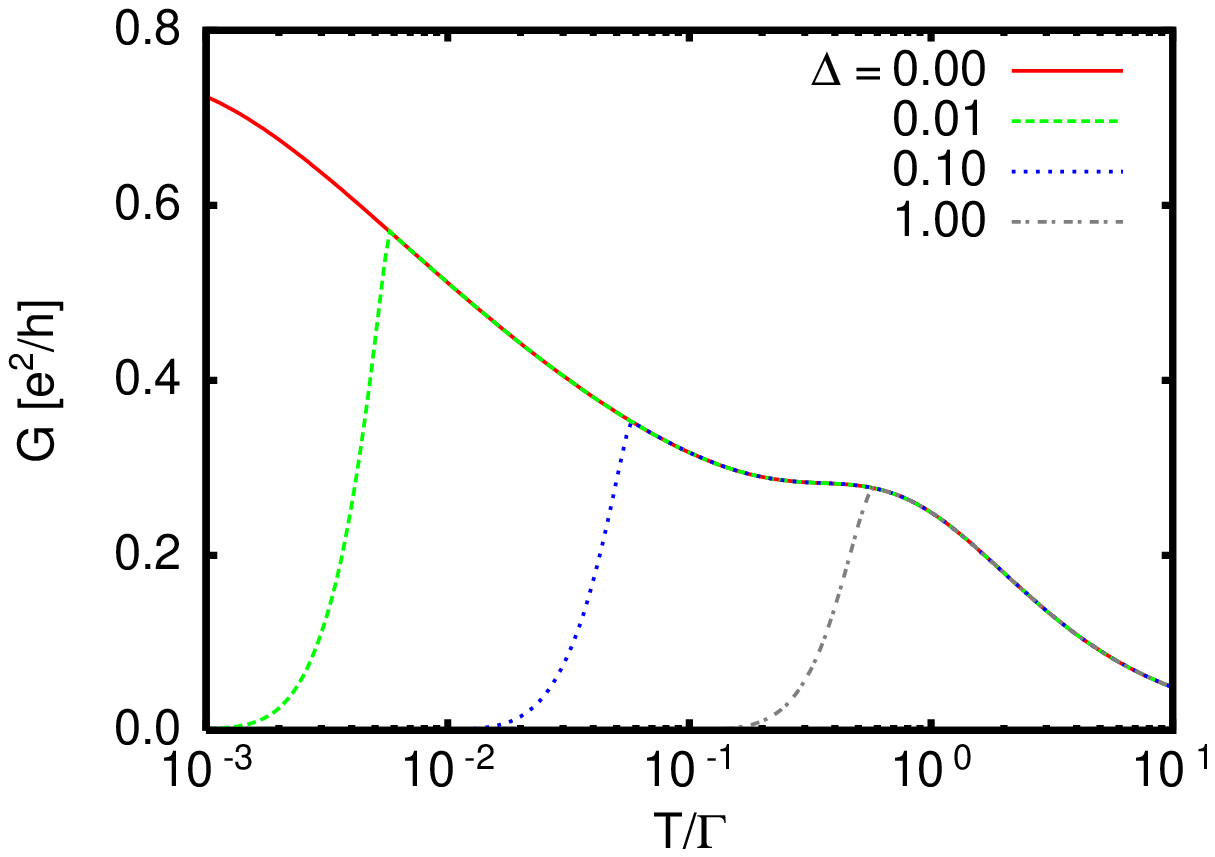}
\includegraphics[width=0.47\textwidth]{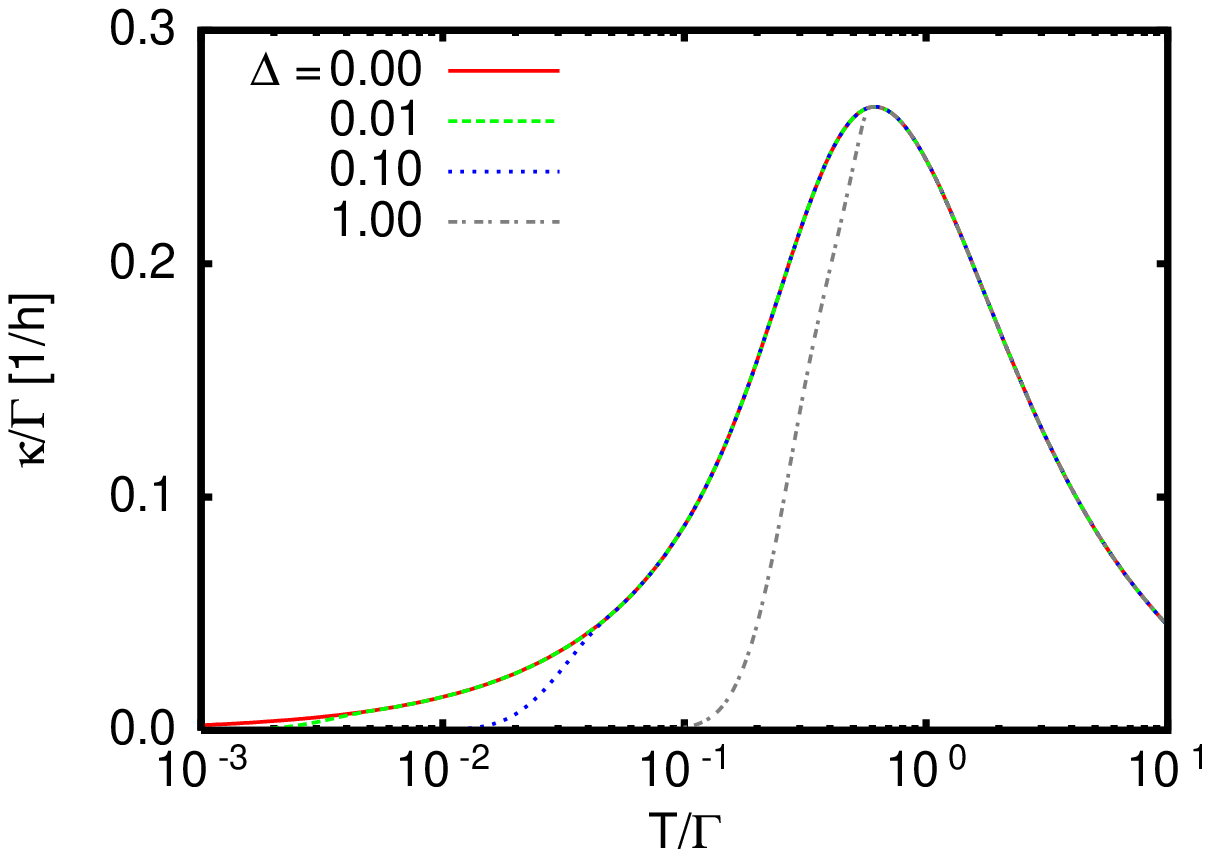}
\caption{\label{Fig1} Temperature dependence of electric (left panel) and 
         thermal conductance (right panel) for various values of 
	 superconducting order parameter. The dot energy level 
	 $\varepsilon_d = -1.75 \Gamma$, and the Kondo temperature 
	 $T_K(\Delta = 0) = 2 \cdot 10^{-2} \Gamma$.}
\end{center}
\end{figure}

Figure \ref{Fig1} shows electric (left panel) and thermal conductance (right
panel) for various values of superconducting order parameter $\Delta$.
One observes strong suppression of the electric ($G$) and thermal ($\kappa$) 
conductance in superconducting state (for $T < T_c = \frac{2 \Delta}{3.52}$). 
The reason for such a behavior is twofold. First, at $T = 0$ the single 
particle transport is completely blocked due to the lack of the electron states 
in superconducting electrode for energies smaller than $\Delta$. Thus the only
contribution to electric transport comes from the Andreev reflections. Note 
that the Andreev reflections do not contribute to the thermal transport (at 
$T = 0$), as the electron in converted into hole, and thus there is no energy 
transfer through the interface. At the same time the Andreev scattering is only 
the mechanism which gives rise to electron transport (the charge of $2e$ is
transferred through the interface). However, strong Coulomb interactions on the 
dot prevent the tunneling of Cooper pairs into QD, and the electrons in each
pair can only tunnel one by one via virtual processes 
\cite{Fazio,Clerk,Cuevas,MK_1}. As a result, the Andreev tunneling is strongly 
suppressed. This is the second reason for such suppression of the electric 
conductance.

\begin{figure}[h]
\begin{center}
\includegraphics[width=0.47\textwidth]{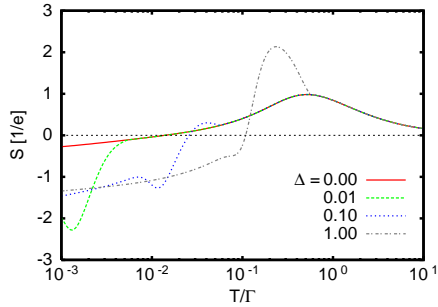}
\caption{\label{Fig2} Linear thermopower $S$ as a function of temperature for
          various values of superconducting order parameter $\Delta$.}
\end{center}
\end{figure}

Corresponding linear thermopower $S$ is shown in Fig. \ref{Fig2}. 
This quantity is very useful measure of the Kondo correlations 
\cite{Boese}-\cite{MK_4}. At high temperatures thermopower is positive 
indicating hole-like transport and at $T \approx \Gamma$ shows a broad maximum 
associated with single particle excitations. At low $T$, where the transport is 
electron-like, the thermopower is negative. In the case of normal leads, the 
thermopower changes sign at the Kondo temperature $T_K$ (solid line in 
Fig. \ref{Fig2}). The sign change can be also understood from the fact that $S$ 
is sensitive to the slope of the density of states (DOS) at the Fermi level. 
With lowering of temperature, the Kondo correlations lead to the development of 
a narrow resonance in the DOS slightly above the Fermi energy, and thus to a 
slope change at the $E_F$. In the presence of superconductivity in one of the 
leads, $S$ changes sign at superconducting transition temperature $T_c$, rather 
than at the Kondo temperature. The sign change of $S$ and its behavior near 
$T_c$ is associated with the shape of superconducting DOS, which is reflected 
in QD density of states. 

\begin{figure}[h]
\begin{center}
\includegraphics[width=0.47\textwidth]{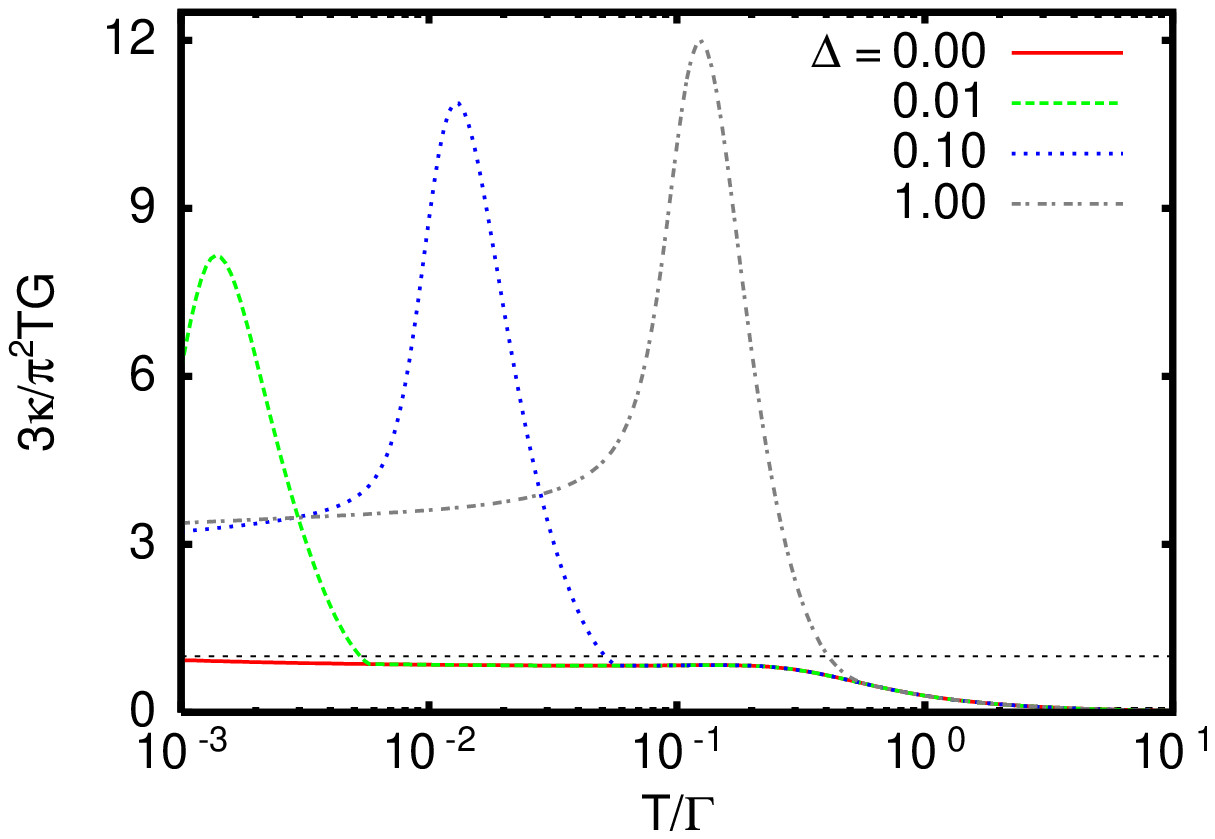}
\includegraphics[width=0.47\textwidth]{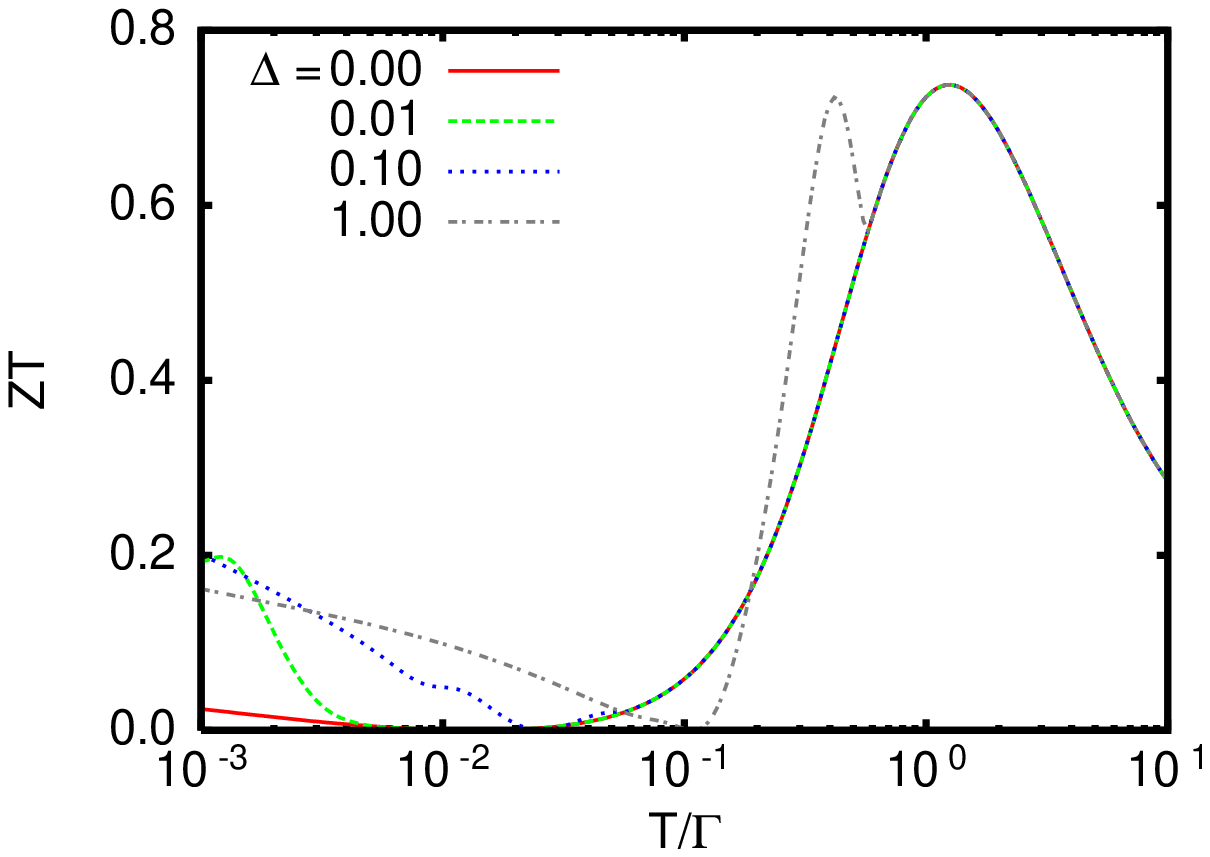}
\caption{\label{Fig3} The Wiedemann-Franz ratio 
         $3 e^2 / \pi^2 (\kappa / T G)$ (left panel) and thermoelectric figure 
	 of merit $ZT = S^2 G T / \kappa$ (right panel) as a function of 
	 temperature $T$.}
\end{center}
\end{figure}

Figure \ref{Fig3} (left panel) shows the temperature dependence of the
Wiedemann-Franz (WF) law which relates thermal and electric conductance via 
the relation $3 e^2/ \pi^2 (\kappa / T G) = 1$. 
This relation describes transport in Fermi liquid bulk metals, and usually is 
violated in nanoscale systems. However, below the Kondo temperature, the WF law 
is recovered, and the system has the Fermi liquid ground state 
\cite{Boese,MK_2,MK_4}. In the high temperature regime, the WF law is violated 
as the transport is due to sequential processes leading to the suppression of 
the thermal transport \cite{MK_2}. This picture is valid for QD coupled to 
normal or to ferromagnetic electrodes. In the presence of superconductivity,
this law is violated even at low temperatures (see left panel of Fig.
\ref{Fig3}), indicating non-Fermi liquid ground state of the system. However,
the superconductivity in one of the electrodes alone is not enough to violate 
the WF law. There must be strong Coulomb repulsion between electrons on the 
dot. In non-interacting QD ($U = 0$) the WF law is still obeyed at low $T$. The 
violation of the WF law in strongly interacting QD has its origin in 
suppression of the Andreev reflections, leading to smaller values of the linear 
conductance $G$ in comparison to non-interacting QD. At the same time, the 
Andreev reflections do not contribute to thermal transport, thus the thermal 
conductance remains unchanged. This explains the violation of the WF law in
strongly interacting N-QD-S system.

Thermoelectric figure of merit $Z = S^2 G / \kappa$ is a direct measure of the
usefulness of the system for applications. For simple systems $Z$ is inversely
proportional to operating temperature, therefore it is convenient to plot 
$ZT$. The right panel of Fig. \ref{Fig3} shows $ZT$ as a function of
temperature. The value of $ZT$ never exceeds 1, which indicates limited 
applicability of the system for thermoelectric power generators or cooling 
systems. Note the enhancement of $ZT$ in superconducting state, which is due to 
the larger values of thermopower in this temperature range 
(see Fig. \ref{Fig2}).

%%%%%%%%%%%%%%%%%%%%%%%%%%%%%%%%%%%%%%%%%%%%%%%%%%%%%%%%%%%%%%%%%%%%%%%%%%%%%%  

\section{\label{conclusions} Summary and conclusions}
In summary, I have studied thermal properties of strongly correlated quantum 
dot coupled to one normal and one superconducting electrode. The 
superconductivity strongly modifies thermal properties of the system. In
particular is responsible for an enhancement of thermopower at temperatures 
close to the superconducting transition temperature. The suppression of Andreev 
reflections, which is due to strong on-dot Coulomb repulsion, leads to a 
violation of the Wiedemann-Franz relation and to non-Fermi liquid ground state.

%%%%%%%%%%%%%%%%%%%%%%%%%%%%%%%%%%%%%%%%%%%%%%%%%%%%%%%%%%%%%%%%%%%%%%%%%%%%%%  

\section*{Acknowledgments}
This work has been supported by the Grant No N N202 1468 33 of the Polish
Ministry of Education and Science.

%%%%%%%%%%%%%%%%%%%%%%%%%%%%%%%%%%%%%%%%%%%%%%%%%%%%%%%%%%%%%%%%%%%%%%%%%%%%%%  

\end{document}